\documentclass[a4paper,notitlepage,twocolumn,footnoteinbib,longbibliography,superscriptaddress,10pt]{revtex4-1}

\usepackage{amsmath,bbold,amssymb,amsthm,mathtools,color,listings,graphicx}
\usepackage[colorlinks=true,linkcolor=blue,citecolor=blue]{hyperref}

\begin{document}

\title{Quantum hyperspins: Highly nonclassical collective behavior in quantum optical parametric oscillators}
\author{Marcello Calvanese Strinati}
\email{marcello.calvanesestrinati@gmail.com}
\affiliation{Centro Ricerche Enrico Fermi (CREF), Via Panisperna 89a, 00184 Rome, Italy}
\author{Claudio Conti}
\affiliation{Physics Department, Sapienza University of Rome, 00185 Rome, Italy}
\affiliation{Centro Ricerche Enrico Fermi (CREF), Via Panisperna 89a, 00184 Rome, Italy}
\date{\today}

\begin{abstract}
We report on the emergence of a highly non-classical collective behavior in quantum parametric oscillators, which we name quantum hyperspin, induced by a tailored nonlinear interaction. This is the second quantized version of classical multidimensional spherical spins, as XY spins in two dimensions, and Heisenberg spins in three dimensions. In the phase space, the quantum hyperspins are represented as spherical shells whose radius scales with the number of particles in a way such that it cannot be factorized even in the limit of large particle number. We show that the nonlinearly coupled quantum oscillators form a high-dimensional entangled state that is surprisingly robust with respect to the coupling with the environment. Such a behavior results from a properly engineered reservoir. Networks of entangled quantum hyperspins are a new approach to quantum simulations for applications in computing, Ising machines, and high-energy physics models. We analyze from first principles through \textit{ab initio} numerical simulations the properties of quantum hyperspins, including the interplay of entanglement and coupling frustration.
\end{abstract}

\maketitle
A paramount challenge in quantum technologies is finding physical systems to be used as qubits~\cite{NielsenChuangBook2016,GZBook2,kim2024}. Ideally, a qubit (i) Must be robust with respect to the interaction with the environment to preserve entanglement, (ii) Must be scalable as quantum advantages are meaningful only at a large scale, and (iii) Must operate at room temperature to reduce operational costs and decoherence. Also, qubits must be coupled beyond local interactions to implement algorithms for combinatorial optimization and tailor quantum simulators~\cite{conti2024quantum}. The most studied technologies, such as superconducting qubits~\cite{wilhelm2008,Gambetta2017,oliver2020}, semiconductors~\cite{10.1063/PT.3.4270,RevModPhys.95.025003}, or trapped ions~\cite{10.1063/1.5088164,RevModPhys.93.025001,Pogorelov2021}, suffer from low temperature operation, limited scalability, and difficulties in controlling couplings~\cite{steuerman2021}.

Room temperature optical qubits have been widely investigated by single-photon sources~\cite{singlephoton} or parametric processes~\cite{PhysRevA.31.2409,science2020}, where the two states of the qubit are encoded using, e.g., different polarization or number ($0$ or $1$ photon) states~\cite{Flamini_2019}. When photon sources are used beyond the single-photon regime, the two logical states can be encoded by two distinct coherent states~\cite{PhysRevA.68.042319,PhysRevLett.100.030503}. This is specifically relevant for non-classical light sources such as degenerate optical parametric oscillators (OPOs). Recently, systems of coupled OPOs forming an Ising machine (IM) found applications in classical coherent computing~\cite{PhysRevA.88.063853,nphoton.2016.68,Hamerly:18}, and their quantum properties are currently under investigation motivated by the perspective to use the quantum nature of parametrically generated light to boost the performance of classical IMs~\cite{s41534-017-0048-9,PhysRevA.102.062419,yamamoto2021nopo}.

\begin{figure*}
\centering
\includegraphics[width=17.5cm]{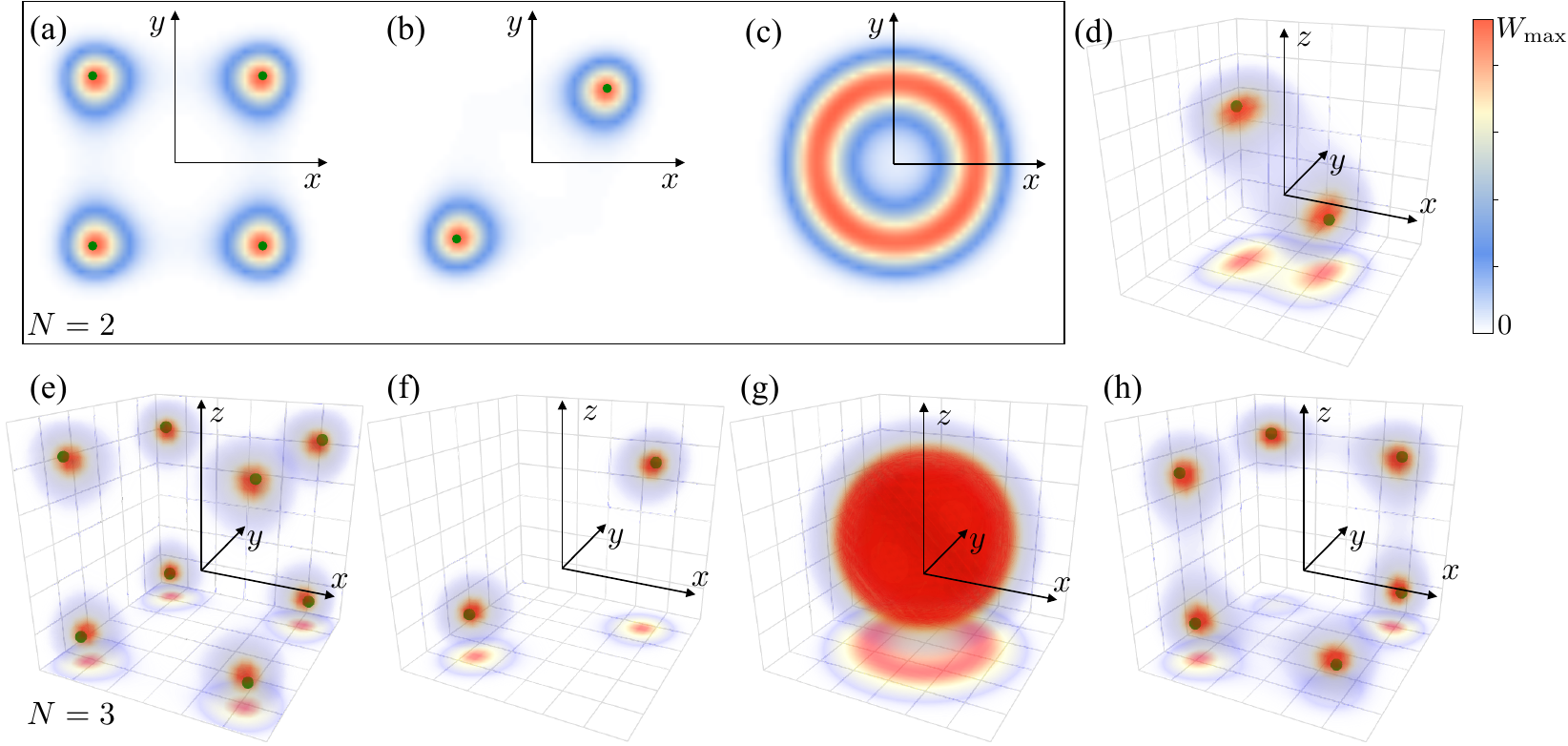}
\caption{Wigner function $W(\vec{\alpha}\,)$ [Eq.~\eqref{eq:hamiltonianlindbladglobalnonlinearity30}] in the real quadrature space for $h$ above threshold. \textbf{(a)}-\textbf{(c)} Colormap of $W(x,y)$ for $N=2$ in the $(x,y)\equiv({\rm Re}[\alpha_1],{\rm Re}[\alpha_2])$ plane. \textbf{(d)}-\textbf{(h)} Volume map of $W(x,y,z)$ for $N=3$ in the $(x,y,z)\equiv({\rm Re}[\alpha_1],{\rm Re}[\alpha_2],{\rm Re}[\alpha_3])$ space with projection on the $xy$-plane. Couplings are: \textbf{(a)},\textbf{(e)} $C_{\mu\nu}=0$ and $W_{\mu\nu}=0$ (decoupled OPOs), \textbf{(b)},\textbf{(f)} $C_{\mu\nu}=c$ and $W_{\mu\nu}=0$ (ferromagnetically coupled OPOs), \textbf{(c)},\textbf{(g)} $C_{\mu\nu}=0$ and $W_{\mu\nu}=\beta$ (XY and Heisenberg hyperspin), \textbf{(d)} $C_{12}=C_{23}=c$ and $C_{13}=-2c$, and \textbf{(h)} $C_{12}=C_{23}=c$ and $C_{13}=-c$, with $W_{\mu\nu}=0$ in both cases (non-Ising and fully-frustrated OPOs~\cite{Calvanese_Strinati_2020}). Numerical parameters: $n_{\rm max}=16$ in \textbf{(g)} and $n_{\rm max}=14$ otherwise, $g=0.5$, $\beta=0.1$, and $c=0.1$. Green dots are the classical fixed points. Color code: From $W=0$ (white) to maximum value $W=W_{\rm max}$ (red) as in the colorbar.}
\label{fig:wignerfunctiotwothreeopos1}
\end{figure*}

However, owing to the coupling with the environment, increasing the pump power to increase the photon number $n$ causes non-classicality to disappear due to decoherence by linear (one-photon) losses~\cite{PhysRevA.104.013715}, eventually switching to a classical behavior~\cite{s41534-017-0048-9}. The phase-space representation of coupled OPOs consists of isolated narrow regions, which move far apart as $n$ increases (Figs.~\ref{fig:wignerfunctiotwothreeopos1} and~\ref{fig:wignerfunctiotwothreeopos2} below), eventually disconnecting. In this situation, the multidimensional Wigner function becomes factorizable, and entanglement is lost~\cite{PhysRevA.102.062419}. One possibility to preserve non-classicality at large photon number is by generating multi-mode Schr\"odinger cat states. They are predicted in Kerr parametric oscillators (KPOs)~\cite{PhysRevA.93.050301,goto2019kpoandopo}, but rely on a negligible coupling with the environment, a condition difficult to realize in practice. One then argues if it is possible to conceive configurations of coupled OPOs with properly engineered dissipation to generate states that preserve entanglement~\cite{PhysRevLett.107.080503,PhysRevResearch.4.013089} at high particle number, to be retained for large-scale computing applications.

In this paper, we propose to engineer the Wigner function in a way that the different regions do not disconnect at any $n$. Specifically, we consider a Wigner function that is a highly-symmetric self-similar surface (a spherical shell), which scales isotropically in any direction when the pump increases. Classically, the spherical shell corresponds to a high-dimensional spherical spin, i.e., a ``hyperspin''. Hyperspins are multidimensional spin variables, as XY spins in two dimensions, or Heisenberg spins in three dimensions~\cite{strinati2022hyperspinmachine}, arising from nonlinearly coupled OPOs, which can be employed to exponentially enhance the performance of classical IMs~\cite{PhysRevLett.132.017301}.

By exploiting an \textit{ab initio} approach with virtually no approximations, we show that a designed engineering of the nonlinear (two-photon) loss in coupled OPOs enables to realize ``quantum hyperspins'', shaping the Wigner function profile as a hyper-spherical shell in phase space. We investigate specifically the cases of two and three coupled OPOs, studying the interplay between frustration and entanglement. At variance with other known Schr\"odinger cat states, quantum hyperspins explicitly exist in the presence of one- and two-photon loss, remarkably preserving entanglement at high photon number.

Our work provides evidence that the nonlinear loss emerging from collective two-photon dissipation can be exploited for a new kind of highly entangled states in OPOs, a strategy so far unexplored. We hence propose hyperspins as novel robust quantum tokens to be employed in computing and fundamental research.

We consider a system of $N$ quantum OPOs subject to parametric drive, and one- and two-photon dissipation processes. For weak pump depletion and large pump mode losses, the system is described by a density operator $\hat\rho$ involving the OPO fields only, whose dynamics obeys the master equation~\cite{PhysRevA.43.6194,PhysRevA.96.053834,goto2019kpoandopo,PhysRevA.102.062419,inuiyamamoto2024,MCS2024}
\begin{equation}
\frac{d}{dt}\hat\rho=\frac{1}{i}\left[\hat H_0,\hat\rho\right]+\mathcal{D}_{\rm 1}(\hat\rho)+\mathcal{D}_{\rm 2}(\hat\rho)\equiv\mathcal{L}\hat\rho\,\, ,
\label{eq:hamiltonianlindbladglobalnonlinearity0}
\end{equation}
where $\mathcal{L}$ is the Liouvillian superoperator. The Hamiltonian $\hat H_0$ describes parametric drive for each OPO:
\begin{equation}
\hat H_0=i\frac{h}{8}\sum_{\mu=1}^{N}\left(\hat a^\dag_\mu\hat a^\dag_\mu-\hat a_\mu\hat a_\mu\right) \,\, ,
\label{eq:hamiltonianlindbladglobalnonlinearity1}
\end{equation}
and $\mathcal{D}_{\rm 1}(\hat\rho)$ and $\mathcal{D}_{\rm 2}(\hat\rho)$ express one- and two-photon dissipation processes, respectively:
\begin{subequations}
\begin{align}
\mathcal{D}_{\rm 1}(\hat\rho)&=\!\sum_{\mu,\nu=1}^{N}\!g_{\mu\nu}\!\left(\hat a_\mu\hat \rho\hat a^\dag_\nu-\frac{1}{2}\left\{\hat a^\dag_\mu\hat a_\nu,\hat \rho\right\}\right)\\
\mathcal{D}_{\rm 2}(\hat\rho)&=\!\frac{1}{2}\sum_{\mu,\nu=1}^{N}\!W_{\mu\nu}\!\left[\hat a^2_\mu\hat \rho{(\hat a^\dag_\nu)}^2\!-\!\frac{1}{2}\left\{{(\hat a^\dag_\mu)}^2 \hat a^2_\nu,\hat \rho\right\}\right] \,.
\end{align}
\label{eq:hamiltonianlindbladglobalnonlinearity2}
\end{subequations}
In Eqs.~\eqref{eq:hamiltonianlindbladglobalnonlinearity1} and~\eqref{eq:hamiltonianlindbladglobalnonlinearity2}, $\hat a_\mu$ ($\hat a^\dag_\mu$) is the photon annihilation (creation) operator for the $\mu$-th OPO, satisfying the commutation relation $[\hat a_\mu,\hat a^\dag_\nu]=\delta_{\mu\nu}$ and $[\hat a_\mu,\hat a^\dag_\nu]=0$, and $h\geq0$ is the pump amplitude. The symmetric matrices $g_{\mu\nu}$ and $W_{\mu\nu}$ describe respectively one- and two-photon dissipation processes: $g_{\mu\mu}\equiv g\geq0$ yields the intrinsic loss for the $\mu$-th OPO and $g_{\mu\nu}\equiv-C_{\mu\nu}$ for $\mu\neq\nu$ quantifies the dissipative linear coupling between the OPOs, with $C_{\mu\mu}=0$. For the two-photon dissipation, $W_{\mu\mu}\equiv\beta$ induces amplitude saturation for each OPO, quantified by the nonlinear coefficient $\beta$, stabilizing the amplitude of the $\mu$-th OPO above threshold, while $W_{\mu\nu}$ for $\mu\neq\nu$ provides the nonlinear coupling between OPOs.

We consider the following cases: (i) For $C_{\mu\nu}\neq0$ and $W_{\mu\nu}=0$, the system consists of $N$ coupled OPOs simulating $N$ coupled Ising spins~\cite{PhysRevA.88.063853}. Instead, (ii) when $C_{\mu\nu}=0$ and $W_{\mu\nu}=\beta$, the system describes a quantum hyperspin formed by $N$ nonlinearly coupled OPOs, with $\mathcal{D}_{\rm 2}(\hat\rho)=\frac{\beta}{2}[\hat L_{\rm hs}\hat\rho\hat L^\dag_{\rm hs}-\frac{1}{2}\{\hat L^\dag_{\rm hs}\hat L_{\rm hs},\hat\rho\}]$ and $\hat L_{\rm hs}=\sum_{\mu=1}^{N}\hat a^2_\mu$. This is seen by the equations of motion for $\hat a_\mu$ from the adjoint master that, in the classical limit, where the operators are replaced by their average values $\hat a_\mu\rightarrow\langle\hat a_\mu\rangle\equiv X_\mu$~\cite{PhysRevA.94.033841}, reduce to the hyperspin machine dynamics of Ref.~\cite{strinati2022hyperspinmachine}, whose real fixed points $(\overline{X}_1,\ldots,\overline{X}_N)$ lie on a hypersphere of radius $S=\sqrt{(h/2-g)/\beta}$, where the overline denotes the steady-state value. Thus, $\overline{X}_\mu/S$ define the $N$ components of a classical hyperspin~\cite{strinati2022hyperspinmachine,footnotesupplementalmaterial}.

We now discuss the numerical simulation of Eq.~\eqref{eq:hamiltonianlindbladglobalnonlinearity0}. Our goal is to obtain the equilibrium density operator $\bar\rho$ from the steady state of Eq.~\eqref{eq:hamiltonianlindbladglobalnonlinearity0}, defined by the condition $\mathcal{L}\overline{\rho}=0$, with subsequent measurement of quantum entanglement. To do so, we extend the \emph{ab initio} method of Ref.~\cite{MCS2024}. We represent quantum operators in the number (Fock) basis of the joint Hilbert space $\mathcal{H}=\bigotimes_{\mu=1}^{N}\mathcal{H}_\mu={\rm span}\{|n_1,\ldots,n_N\rangle\}$, where $\mathcal{H}_\mu={\rm span}\{|n_\mu\rangle\}_{n_\mu=0}^{\infty}$ is the local Hilbert space of the $\mu$-th OPO, and $|n_1,\ldots,n_N\rangle\equiv\bigotimes_{\mu=1}^{N}|n_\mu\rangle$ is the joint Fock state of $n_\mu$ photons in each individual OPO. Consequently, $\hat\rho$ is mapped onto a $2N$-index tensor $\rho_{m_1,\ldots,m_N;n_1,\ldots,n_N}\coloneqq\langle m_1,\ldots,m_N|\hat\rho|n_1,\ldots,n_N\rangle$, and the Liouvillian supeoperator is represented by the sparse $4N$-index tensor $\mathcal{L}^{r_1,\ldots,r_N;s_1,\ldots,s_N}_{m_1,\ldots,m_N;n_1,\ldots,n_N}$ encoding the nonzero elements of the right-hand side of Eq.~\eqref{eq:hamiltonianlindbladglobalnonlinearity0} projected onto the Fock basis. Then, $\overline{\rho}$ is determined as the eigenvector of $\mathcal{L}$ corresponding to the zero eigenvalue~\cite{PhysRevA.98.042118,footnotesupplementalmaterial}.

While in general the number states are upper unbounded ($n_\mu=0,\ldots,\infty$), and thus $\hat\rho$ is represented as an infinite-dimensional tensor, the presence of pump-saturation effect in Eq.~\eqref{eq:hamiltonianlindbladglobalnonlinearity2} naturally fixes an upper bound to the actual average number of photons $\overline{n}_{\mu}$ in the steady state, by stabilizing the amplitude dynamics. This fact allows us to truncate the local Hilbert spaces up to a maximum photon number $n_{\rm max}$ such that $\bar{\rho}_{m_1,\ldots,m_N;n_1,\ldots,n_N}\simeq0$ for all indexes larger than $n_{\rm max}$. This ensures that any observable computed from $\hat{\rho}$ is unaffected by the truncation of the Hilbert space.

\begin{figure}[t]
\centering
\includegraphics[width=8.5cm]{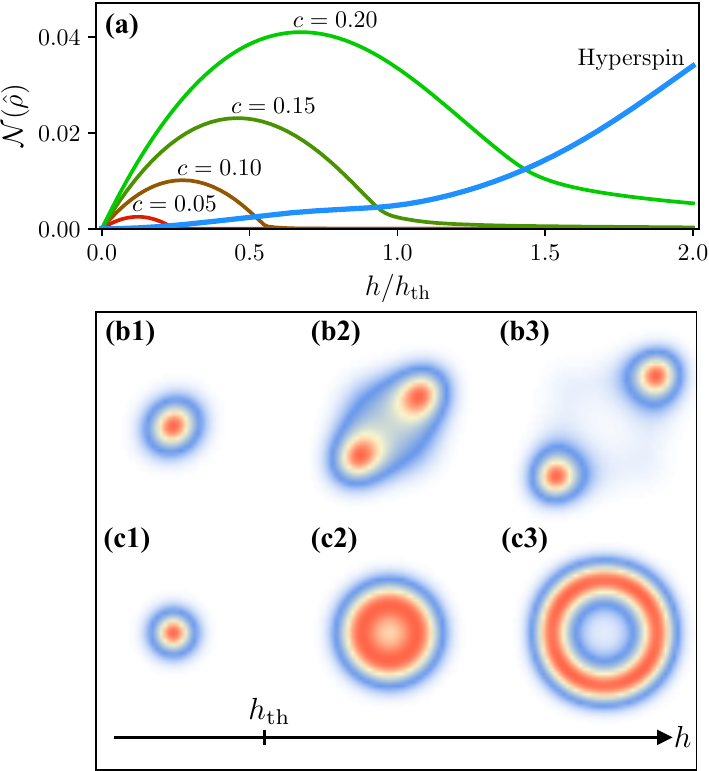}
\caption{\textbf{(a)} Entanglement negativity for $N=2$ as a function of pump relative to threshold. Thin lines show $\mathcal{N}(\hat\rho)$ for ferromagnetically coupled OPOs with coupling $c$ from $0.05$ (red) to $0.2$ (green) as in the labels. Blue tick line shows $\mathcal{N}(\hat\rho)$ for the XY hyperspin. For linearly coupled OPOs, $\mathcal{N}(\hat\rho)$ is non-monotonic, where both the maximum entanglement and the pump value at which the maximum is found increase with $c$. Instead, the hyperspin shows a monotonic increase of entanglement with increasing pump. \textbf{(b1)}-\textbf{(b3)} Colormap of $W(\vec{\alpha}\,)$ for ferromagnetically coupled OPOs, and \textbf{(c1)}-\textbf{(c3)} for one XY hyperspin (see Fig.~\ref{fig:wignerfunctiotwothreeopos1}), for pump increasing from below to above threshold (from left to right panels). Differently from two coupled OPOs, $W(\vec{\alpha}\,)$ for the hyperspin remains connected (specifically, self-similar) at any pump.}
\label{fig:wignerfunctiotwothreeopos2}
\end{figure}

\begin{figure*}[t]
\centering
\includegraphics[width=17.5cm]{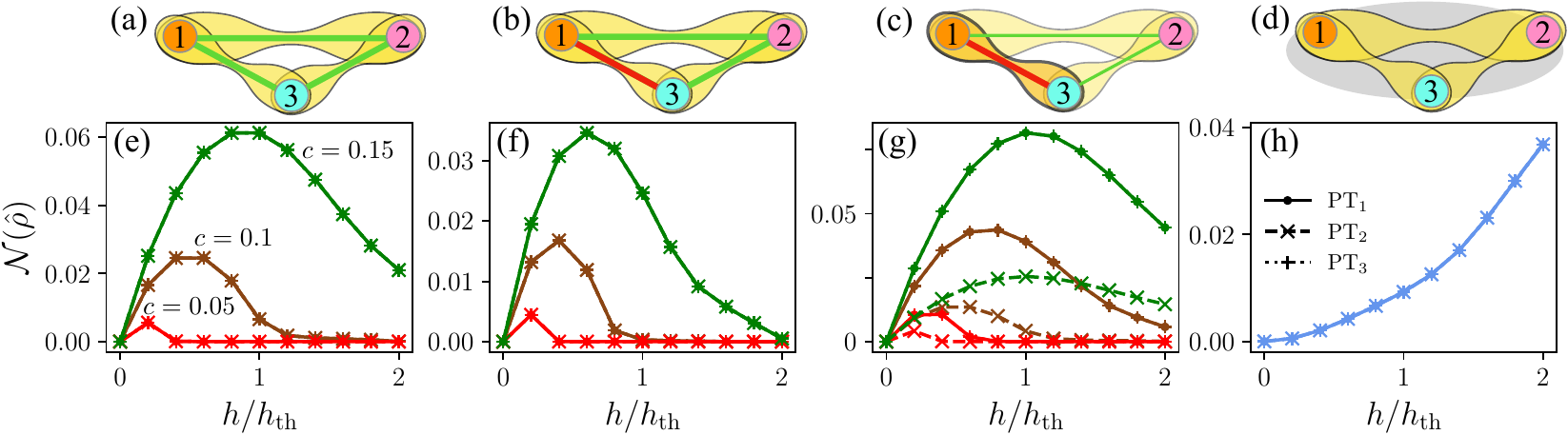}
\caption{\textbf{(a)}-\textbf{(d)} Sketch of coupling and entanglement for $N=3$ \textbf{(a)} ferromagnetically coupled OPOs, \textbf{(b)} fully-frustrated OPOs, \textbf{(c)} non-Ising graph, and \textbf{(d)} Heisenberg hyperspin (Fig.~\ref{fig:wignerfunctiotwothreeopos1}). Colored dots depict the OPOs, green and red lines their positive and negative coupling, respectively, and yellow connections depict mutual entanglement. The gray circle in panel \textbf{(d)} depicts the nonlinear coupling. \textbf{(e)}-\textbf{(h)} Entanglement negativity as a function of pump relative to threshold for the corresponding connectivity in \textbf{(a)}-\textbf{(d)}. Panels \textbf{(e)}-\textbf{(g)} show $\mathcal{N}(\hat\rho)$ for $c=0.05$ (red), $c=0.1$ (brown), and $c=0.15$ (green), while panel \textbf{(h)} shows $\mathcal{N}(\hat\rho)$ for the hyperspin. For each $c$ and for the hyperspin, $\mathcal{N}(\hat\rho)$ is shown from three different partial transpositions: ${\rm PT}_1$ (dots with solid line), ${\rm PT}_2$ (crosses ``x'' with dashed line), and ${\rm PT}_3$ (pluses ``+'' with dash-dotted line). Similar to Fig.~\ref{fig:wignerfunctiotwothreeopos2}, $\mathcal{N}(\hat\rho)$ is non-monotonic with $h$ for linearly coupled OPOs, with peak position and value increasing with $c$, while it monotonically increases for the hyperspin. Different partial transpositions yield the same $\mathcal{N}(\hat\rho)$ in all cases except for \textbf{(g)}, where ${\rm PT}_2$ yields the smallest value of entanglement, while ${\rm PT}_1$ and ${\rm PT}_3$ yield the same. Other numerical parameters are: $g=0.5$, $\beta=0.1$, and $n_{\rm max}=12$ apart from the points in panel \textbf{(h)} for $h/h_{\rm th}>1$ where $n_{\rm max}=16$ was used.}
\label{fig:wignerfunctiotwothreeopos3}
\end{figure*}

We remark that the advantage of this \emph{ab initio} method is that it gives access to the \emph{full} density matrix, which is particularly useful to \emph{measure} quantum entanglement in the state of coupled OPOs, without approximation. Clearly, the choice of number states requires in general the numerical determination of a large number of parameters, which scales as $n_{\rm max}^{2N}$ for $N$ OPOs. For a too large number of photons $n_{\rm max}$ or large system size $N$, the increasing numerical complexity makes the use alternative approaches like phase-space methods as the Wigner or coherent state $P$-representation~\cite{PhysRevA.43.6194,Kiesewetter:22} necessary.

The quantum state in phase space is visualized via the Wigner function $W(\vec{\alpha}\,)$~\cite{PhysRev.177.1857,PhysRev.177.1882}. For a $N$-mode state with complex quadratures $\vec{\alpha}\equiv(\alpha_1,\ldots,\alpha_N)$, this is defined as
\begin{equation}
W\left(\vec{\alpha}\,\right)={\left(\frac{2}{\pi}\right)}^N\!\!{\rm Tr}\left[\hat\rho\,\prod_{\mu=1}^{N}\hat D_{\mu,2\alpha_\mu}\,\prod_{\nu=1}^{N}e^{i\pi\hat a^\dag_\nu\hat a_\nu}\right] \,,
\label{eq:hamiltonianlindbladglobalnonlinearity30}
\end{equation}
where $\hat D_{\mu,\alpha_\mu}\coloneqq e^{\alpha_\mu\hat a^\dag_\mu-\alpha^*_\mu\hat a_\mu}$ is the displacement operator for the $\mu$-th OPO. The Wigner function in Eq.~\eqref{eq:hamiltonianlindbladglobalnonlinearity30} is a real function of $\alpha_\mu\in\mathbb{C}$, where the real and imaginary part of $\alpha_\mu$ represent, respectively, the position and momentum quadrature of the $\mu$-th OPO, whose average values approximately correspond to the real and imaginary part of the classical amplitude $X_\mu$.

We compare in Fig.~\ref{fig:wignerfunctiotwothreeopos1} the numerically determined Wigner functions from Eq.~\eqref{eq:hamiltonianlindbladglobalnonlinearity30} for $N=2,3$ and real $\alpha_\mu$, for $h$ high above the classical oscillation threshold $h_{\rm th}=2(g-\lambda_{\rm max})$, where $\lambda_{\rm max}$ is the largest eigenvalue of $C_{\mu\nu}$, and different connectivities: \textbf{(a)}-\textbf{(c)} and \textbf{(e)}-\textbf{(g)} show $W(\vec{\alpha}\,)$ for decoupled OPOs, ferromagnetically coupled OPOs, and the hyperspin, for $N=2$ and $N=3$, respectively. For decoupled and ferromagnetically coupled OPOs, the Wigner function consists respectively of $2^N$ and $2$ lobes, whose position is consistent with the classical fixed points (green dots), encoding different Ising solutions. Instead, for the hyperspin, the Wigner function displays a doughnut and shell shape for $N=2$ and $N=3$, respectively, indicating that the system most probably occupies a $N$-dimensional hyper-spherical shell in quadrature space whose mean radius is approximately $S$ found in the classical limit.

Panels \textbf{(d)},\textbf{(h)} refer to $N=3$ with coupling as in Ref.~\cite{Calvanese_Strinati_2020}, corresponding to two different frustrated connectivities: Degenerate and zero (or ``non-Ising'') coupling~\cite{PhysRevLett.126.143901}, respectively. In the first case, frustration gives rise to six degenerate solutions of the simulated Ising model, while in the second case, it causes OPO $2$ to be at zero amplitude (characterized by a Wigner function centered about the $xz$-plane), thus preventing the definition of valid Ising solution. We stress that the Wigner function in the coupled-OPO and hyperspin case is profoundly different: In the first case, increasing pump drives the system towards a configuration of \emph{disconnected} lobes, with exponentially suppressed overlap, while in the second case, the Wigner function remains \emph{fully connected} at any pump. 

Next, we study the presence of quantum entanglement in the system, for the connectivities in Fig.~\ref{fig:wignerfunctiotwothreeopos1}. A notable advantage of the number-state expansion of $\hat\rho$ is that it straightforwardly allows the quantification of quantum entanglement following the partial transpose (PT) criterion~\cite{PhysRevLett.77.1413,HORODECKI19961}, which states that if $\hat\rho$ is separable, then the partially transposed density operators $\mathrm{PT}_\nu\hat\rho$ necessarily represent valid states (i.e., with unit trace and positive), with ${\rm PT}_\nu\coloneqq(\bigotimes_{\mu=1}^{\nu-1}\mathbb{1}_\mu)\hat T_\nu(\bigotimes_{\mu=\nu+1}^{N}\mathbb{1}_\mu)$, where $\mathbb{1}_\mu$ and $\hat{T}_\nu$ are the identity and transposition operators on $\mathcal{H}_\mu$ and $\mathcal{H}_\nu$, respectively. The violation of such criterion is then a sufficient condition for entanglement. From our simulation, we can readily detect the presence of entanglement by checking that at least one eigenvalue of the PT density matrix $\mathrm{PT}_\nu\hat\rho$, obtained from $\overline{\rho}_{m_1,\ldots,m_N;n_1,\ldots,n_N}$ by swapping the indexes $m_\nu$ and $n_\nu$, is negative. Furthermore, if $\{\theta_k\}$ are the negative eigenvalues of $\mathrm{PT}_\nu\hat\rho$, we measure the amount of entanglement by the entanglement negativity $\mathcal{N}(\hat\rho)\coloneqq\sum_{k}|\theta_k|$ for the selected transposition~\cite{PhysRevA.58.883,PhysRevA.65.032314}.

Figure~\ref{fig:wignerfunctiotwothreeopos2}\textbf{a} shows $\mathcal{N}(\hat\rho)$ as a function of $h/h_{\rm th}$ across the threshold for $N=2$, which is the same for both ${\rm PT}_1$ and ${\rm PT}_2$ since the density matrix is real and symmetric. Solid lines refer to the negativity of two ferromagnetically coupled OPOs in Fig.~\ref{fig:wignerfunctiotwothreeopos1}\textbf{b} with $c$ as in the labels, while ``Hyperspin'' refers to the XY hyperspin in Fig.~\ref{fig:wignerfunctiotwothreeopos1}\textbf{c}. As evident, $\mathcal{N}(\hat\rho)$ has a non-monotonic behaviour, with both peak position and value increasing with $c$. In other words, the maximum entanglement is found at an optimal value of pump amplitude relative to $h_{\rm th}$ that depends on the coupling strength. Importantly, this optimal pump value generically does not coincide with the oscillation threshold value. For large pump, $\mathcal{N}(\hat\rho)$ decreases signaling the tendency of the system to become separable. This fact is manifest when looking at the Wigner function in panels \textbf{(b)} as the pump is increased from below threshold [panel \textbf{(b1)}] to well above threshold [panel \textbf{(b3)}].

A drastically different scenario is found for the XY hyperspin: We find that, while $\mathcal{N}(\hat\rho)$ takes lower values around the threshold compared the two linearly coupled OPOs, it monotonically grows with increasing pump, suggesting the emergence of a highly entangled state, which does not separate at high pump. Accordingly, $W(\vec{\alpha}\,)$ in panels \textbf{(c)} preserves its spherical-shell shape at any pump, therefore remaining always connected (in particular, self-similar). This result has a remarkable implication: Differently from linearly coupled OPOs, increasing pump does not drive the quantum hyperspin towards its classical limit.

The entanglement analysis for $N=3$ is reported in Fig.~\ref{fig:wignerfunctiotwothreeopos3}, for linearly coupled OPOs with connectivities as in Fig.~\ref{fig:wignerfunctiotwothreeopos1}\textbf{f},\textbf{h},\textbf{d} in panels \textbf{(e)}-\textbf{(g)}, respectively, and for the Heisenberg hyperspin in panel \textbf{(h)}. Due to the presence of different coupling constants in $C_{\mu\nu}$, the entanglement negativity generically depends on the specific partial transposition. Thus, we show $\mathcal{N}(\hat\rho)$ for ${\rm PT}_\nu$ with $\nu=1,2,3$ (other partial transpositions reduce to these three) for each connectivity and $c$. The picture that emerges is consistent with Fig.~\ref{fig:wignerfunctiotwothreeopos2}: For three linearly coupled OPOs, $\mathcal{N}(\hat\rho)$ for all $c$ and ${\rm PT}_\nu$ displays a non-monotonic behaviour, where both the optimal value of $h$ at which entanglement is maximum and the corresponding maximum value increase with $c$. On the contrary, the Heisenberg hyperspin displays a monotonic growth of entanglement. The fact that $\mathcal{N}(\hat\rho)$ for all ${\rm PT}_\nu$ is the same in panels \textbf{(e)},\textbf{(f)}, and \textbf{(h)} is a consequence of the fact that all OPOs for those connectivities share the same coupling weight. This is instead not found in panel \textbf{(g)}, where ${\rm PT}_2$ yields a smaller $\mathcal{N}(\hat\rho)$ compared to ${\rm PT}_1$ and ${\rm PT}_3$ due to the overall smallest coupling of OPO $2$ compared to the other two OPOs.

In conclusion, we reported on a detailed, \emph{ab initio} study of entanglement in two and three OPOs coupled with both linear and nonlinear coupling, arising from nonlocal one- and two-photon dissipation, respectively. Following the partial transpose criterion, we quantified entanglement by the entanglement negativity from the density matrix obtained by exact diagonalization of the Liouvillian superoperator in the Fock (number) basis. We found that, for zero nonlinear coupling and different linear coupling connectivities (including frustrated ones), entanglement shows a non-monotonic behaviour for increasing pump. The maximum entanglement is found at an optimal pump, which depends on the coupling and generically does not coincide with the oscillation threshold. In contrast, the quantum hyperspin obtained by coupling OPOs with purely nonlinear coupling showed a monotonically growing entanglement with increasing pump, unveiling the emergence of a highly-entangled state with no classical counterpart in the large-pump regime.

Therefore, a multidimensional oscillator with nonlinearly coupled degrees of freedom acts as a kind of high-dimensional quantum spin,
whose phase-space representation is non-factorizable even at large particle number. Besides OPOs, KPOs have been proposed to generate macroscopically entangled states~\cite{goto2019kpoandopo}. A natural question following our work is whether suitable setups of KPOs can realize quantum hyperspins, and importantly, how the interplay of linear and nonlinear couplings affect quantum correlations. Hyperspins also emerge in models for quantum chromodynamics and high energy physics. Realizing devices that support hyperspins with variable topologies and configurations may open the way to new classes of quantum simulators and unveil an unexplored and rich physics. Importantly, classical hyperspins boost combinatorial optimization~\cite{PhysRevLett.132.017301}, and whether their quantum counterpart sustains a further quantum advantage is an open question, which we will address in future work.

\vspace{0.2cm}
We acknowledge support from HORIZON-ERC-2023-ADG HYPERSPIM project grant number 101139828.


%

\end{document}